\documentclass{ieeeaccess}

\usepackage[T1]{fontenc}
\usepackage{lmodern}

\usepackage{cite}
\usepackage{amsmath,amssymb,amsfonts}
\usepackage{algorithmic}
\usepackage{algorithm} 
\usepackage{graphicx}
\usepackage{textcomp}
\usepackage{xcolor}
\usepackage{braket}
\usepackage{booktabs}
\usepackage{pdfpages}
\usepackage{subcaption}
\usepackage{url}
\usepackage[hidelinks]{hyperref}

\def\BibTeX{{\rm B\kern-.05em{\sc i\kern-.025em b}\kern-.08em
    T\kern-.1667em\lower.7ex\hbox{E}\kern-.125emX}}

\begin{document}
\history{Date of publication xxxx 00, 0000, date of current version xxxx 00, 0000.}
\doi{10.1109/TQE.202X.DOI}

\title{Hardware-Aware Quantum Support Vector Machines}

\author{\uppercase{Adil Mubashir Chaudhry}\authorrefmark{1},
\uppercase{Ali Raza Haider}\authorrefmark{1},
\uppercase{Hanzla Khan}\authorrefmark{1}, and
\uppercase{Dr. Muhammad Faryad}\authorrefmark{1}, \IEEEmembership{Member, IEEE}}

\address[1]{School of Science and Engineering, Lahore University of Management Sciences, Lahore, Pakistan (email: 24280007@lums.edu.pk, 24280070@lums.edu.pk, 24280020@lums.edu.pk, muhammad.faryad@lums.edu.pk)}

\markboth
{Chaudhry \headeretal: Hardware-Aware Quantum Support Vector Machines}
{Chaudhry \headeretal: Hardware-Aware Quantum Support Vector Machines}

\corresp{Corresponding author: Dr. Muhammad Faryad (email: muhammad.faryad@lums.edu.pk).}

\begin{abstract}
Deploying quantum machine learning algorithms on near-term quantum hardware requires circuits that respect device-specific gate sets, connectivity constraints, and noise characteristics. We present a hardware-aware Neural Architecture Search (NAS) approach for designing quantum feature maps that are natively executable on IBM quantum processors without transpilation overhead. Using genetic algorithms to evolve circuit architectures constrained to IBM Torino native gates (ECR, RZ, SX, X), we demonstrate that automated architecture search can discover quantum Support Vector Machine (QSVM) feature maps achieving competitive performance while guaranteeing hardware compatibility. Evaluated on the UCI Breast Cancer Wisconsin dataset, our hardware-aware NAS discovers a 12-gate circuit using exclusively IBM native gates (6$\times$ECR, 3$\times$SX, 3$\times$RZ) that achieves 91.23\% accuracy on 10 qubits—matching unconstrained gate search while requiring zero transpilation. This represents a 27 percentage point improvement over hand-crafted quantum feature maps (64\% accuracy) and approaches classical RBF SVM baseline (93\%). We show that removing architectural constraints (fixed RZ placement) within hardware-aware search yields 3.5 percentage point gains, and that 100\% native gate usage eliminates decomposition errors that plague universal gate compilations. Our work demonstrates that hardware-aware NAS makes quantum kernel methods practically deployable on current noisy intermediate-scale quantum (NISQ) devices, with circuit architectures ready for immediate execution without modification.
\end{abstract}


\titlepgskip=-15pt

\maketitle

\section{Introduction}

Quantum Support Vector Machines (QSVMs) leverage quantum feature maps to embed classical data into exponentially large Hilbert spaces, potentially enabling pattern recognition capabilities beyond those achievable by classical kernels \cite{havlicek2019supervised,schuld2019quantum}. Despite this theoretical promise, designing effective quantum feature maps has proven challenging in practice. Hand-crafted circuits often fail to capture dataset structure, while directly optimizing continuous parameters is susceptible to barren plateaus and becomes computationally expensive \cite{schuld2021effect}. These limitations have motivated alternative approaches that search over circuit architectures rather than tuning parameters.

In this work, we address the practical deployment challenge of quantum feature maps through hardware-aware Neural Architecture Search (NAS), implemented using genetic algorithms that evolve quantum circuits directly \cite{du2022quantum}. Rather than designing universal circuits requiring transpilation, our approach constrains the search space to IBM Torino native gates (ECR, RZ, SX, X) and respects device connectivity from the outset. This ensures discovered circuits execute efficiently on actual quantum hardware without gate decomposition, SWAP insertion, or other compilation overhead that degrades fidelity. We compare hardware-aware NAS against classical SVM baselines, hand-crafted quantum feature maps (ZZ, Pauli), unconstrained all-gates NAS, noise-aware variants, and sparsity-constrained search. Our results demonstrate that hardware-aware NAS discovers compact, natively executable quantum feature maps that substantially outperform hand-crafted alternatives while maintaining practical deployability on near-term devices.

The effectiveness of QSVMs and quantum kernels has been explored extensively in the literature. Schuld and Killoran \cite{schuld2019quantum} formalized quantum feature maps as implicit kernel methods, and Havlíček et al. \cite{havlicek2019supervised} demonstrated one of the earliest practical QSVM implementations. Work on kernel evaluation quality, such as kernel-target alignment introduced by Cristianini et al. \cite{cristianini2001kernel} and later centered alignment by Cortes et al. \cite{cortes2012algorithms}, has influenced approaches for assessing and training quantum kernels. Hubregtsen et al. \cite{hubregtsen2022training} applied alignment concepts to quantum circuits, showing that alignment-based optimization can improve QSVM performance.

Recent findings also highlight fundamental challenges in quantum kernel design. Kübler et al. \cite{kubler2021inductive} analyzed the inductive bias introduced by quantum kernels and showed that excessive expressibility can harm generalization, while Schuld et al. \cite{schuld2021effect} emphasized that data encoding strategies are crucial for ensuring meaningful kernel structure. These insights reinforce the need to explore architectures systematically rather than relying on hand-crafted or overly expressive circuits.

Neural Architecture Search has therefore emerged as a promising tool for generating effective quantum circuits. Prior work has applied evolutionary algorithms \cite{skolik2022quantum}, reinforcement learning \cite{yao2021policy}, and differentiable architecture methods \cite{du2022quantum} to quantum circuit design. Our work extends this line of research to quantum kernel learning by searching over gate vocabularies, connectivity, and entanglement structure, enabling the discovery of hardware-efficient and noise-robust feature maps.

Our results empirically validate three key hypotheses: first, that hardware-aware constraints do not prohibitively restrict performance—constrained NAS matches unconstrained search; second, that automated architecture search substantially outperforms manual circuit design even under strict hardware limitations; and third, that native gate execution without transpilation provides a viable path toward practical quantum kernel deployment on NISQ devices.

\section{Methods}

\subsection{Dataset and Preprocessing}

We use the UCI Breast Cancer Wisconsin (Diagnostic) dataset, a binary classification benchmark with 569 samples and 30 real-valued features. The data is split into 80\% training and 20\% test sets using stratified sampling. All features are standardized using z-score normalization. To make quantum kernel computation tractable, we select the ten highest-variance features and reduce the dataset to a six-qubit representation for initial experiments. This preserves the dominant structure of the dataset while reducing circuit and kernel evaluation cost.

During architecture search, computing full quantum kernels for every candidate feature map is expensive, so we further subsample 200 training points for NAS evaluation. This strikes a balance between computational feasibility and robustness of fitness estimation.

\subsection{Classical SVM Baselines}

We evaluate two classical models as reference points. A linear SVM is trained with the regularization parameter tuned by cross-validation over five orders of magnitude. An RBF SVM is tuned by grid search over both $C$ and $\gamma$. Both baselines are evaluated on the held-out test set using accuracy, precision, recall, and F1-score.

\subsection{Quantum Feature Maps}

We evaluate several standard quantum feature maps provided by Qiskit, including the ZZFeatureMap, Pauli-based encodings, ZFeatureMap, RawFeatureVector, and efficient rotation-entangler constructions. Each feature map is used to generate a fidelity-based quantum kernel computed through the FidelityQuantumKernel class. A classical SVM with a precomputed kernel is then trained on the resulting Gram matrix and evaluated on the test set.

\subsection{Genetic Algorithm for Architecture Search}

To search over quantum feature-map architectures, we implement a genetic algorithm (GA) that directly operates on discrete circuit structures. Each candidate circuit is represented as a genome: a variable-length list of gate tokens drawn from the allowed vocabulary. For hardware-aware experiments, this vocabulary includes only IBM Torino native gates (RZ$_q$, SX$_q$, X$_q$, and ECR$_{c,t}$), derived from the device's coupling map. Genome lengths are restricted to 4--12 gates.

A genome is converted into a parameterized feature map by first applying data-dependent RZ($x_i$) rotations to all qubits, followed by the sequence of gates encoded in the genome. The fidelity-based quantum kernel is then computed with Qiskit's FidelityQuantumKernel, and the genome's fitness is defined as cross-validated QSVM accuracy on a 200-sample subsample of the training set.

The GA initializes a population of eight random genomes and evolves them over four generations. In each generation, genomes are ranked by fitness and the top two are preserved via elitism. New genomes are produced via tournament selection ($k=3$), single-point crossover, and mutation. Mutation is implemented with each gene replaced with probability $p_{mut}=0.25$, with additional insertion and deletion events occurring with probability 0.1. This introduces sufficient variability for exploring the architecture space.

The evolutionary loop continues until all generations are completed, at which point the genome with highest QSVM accuracy is selected as the final feature-map architecture.

\begin{algorithm}[t]
\caption{Hardware-Aware Genetic Search for Quantum Feature Maps}
\label{alg:nas}
\begin{algorithmic}[1]
\REQUIRE Training data $(X, y)$, gate set $G$, coupling map, population size $P$, generations $T$
\STATE Standardize features and select top-variance subset
\STATE Subsample 200 training samples for NAS evaluation
\STATE Initialize population $\mathcal{P} = \{g_1, \ldots, g_P\}$ with random genomes (length 4--12)
\FOR{$t = 1$ to $T$}
    \FOR{each genome $g$ in $\mathcal{P}$}
        \STATE Build quantum feature map $U_g$:
        \STATE \quad Apply base RZ($x_i$) encoding
        \STATE \quad Append gates in genome sequence
        \STATE Compute quantum kernel $K_g = \text{FidelityQuantumKernel}(U_g)$
        \STATE Evaluate fitness using cross-validated QSVM accuracy
    \ENDFOR
    \STATE Rank genomes by fitness
    \STATE Select top two genomes for elitism
    \STATE Initialize new population $\mathcal{P}'$
    \STATE Add elite genomes to $\mathcal{P}'$
    \WHILE{$|\mathcal{P}'| < P$}
        \STATE Select parents via tournament selection ($k=3$)
        \STATE Apply single-point crossover to produce children
        \STATE Mutate each genome with probability $p_{mut} \approx 0.25$
        \STATE Add children to $\mathcal{P}'$
    \ENDWHILE
    \STATE $\mathcal{P} \leftarrow \mathcal{P}'$
\ENDFOR
\RETURN Best genome $g^*$ and corresponding circuit $U_{g^*}$
\end{algorithmic}
\end{algorithm}

\subsection{NAS Variants}

We explore multiple NAS configurations to understand the trade-offs between expressiveness, hardware compatibility, and robustness:

\textbf{IBM Hardware-Aware NAS (Fixed RZ):} The search space is restricted to the native gate set of IBM Torino, consisting of RZ, SX, X, and ECR gates, limited by its heavy-hex connectivity. The circuit includes mandatory initial RZ rotations for data encoding.

\textbf{IBM Hardware-Aware NAS (No Fixed RZ):} Similar to the above but removes the mandatory initial RZ constraint, allowing NAS to discover optimal rotation placements.

\textbf{Unconstrained All-Gates NAS:} The search spans a larger gate set including RX, RY, H, S, CZ, and CX, allowing more expressive but less hardware-efficient circuits.

\textbf{Noise-Resilient NAS:} Incorporates depolarizing and amplitude damping noise when evaluating quantum kernels, encouraging discovery of robust circuit architectures.

\textbf{Sparsity-Constrained NAS:} Penalizes excessive use of two-qubit entanglers to bias the search toward shallow architectures.

\section{Results}

\subsection{Classical Baselines}

Table~\ref{tab:classical} summarizes classical SVM performance on the test set. The RBF SVM achieves 93.0\% accuracy with precision 93.2\%, recall 95.8\%, and F1-score 94.5\%, establishing a strong baseline. Linear SVM achieves 91.2\% accuracy, demonstrating that classical methods remain highly competitive on this dataset.

\begin{table}[ht]
\centering
\caption{Classical SVM performance on the Breast Cancer Wisconsin test set.}
\label{tab:classical}
\begin{tabular}{lcccc}
\hline
Model & Accuracy & Precision & Recall & F1 \\
\hline
Linear SVM & 0.912 & 0.943 & 0.917 & 0.930 \\
RBF SVM & 0.930 & 0.932 & 0.958 & 0.945 \\
\hline
\end{tabular}
\end{table}

\subsection{Hand-Crafted Quantum Feature Maps}

Table~\ref{tab:handcrafted} presents QSVM results using standard feature maps. Standard entangling feature maps (ZZ, Pauli) perform poorly, achieving only 63--64\% accuracy, while amplitude-based encoding (RawFeatureVector) matches classical RBF SVM performance. This demonstrates how manual circuit design often fails to capture the underlying structure of the dataset.

\begin{table}[ht]
\centering
\caption{QSVM performance with hand-crafted quantum feature maps on the test set.}
\label{tab:handcrafted}
\begin{tabular}{lcccc}
\hline
Feature Map & Accuracy & Precision & Recall & F1 \\
\hline
ZZFeatureMap & 0.640 & 0.637 & 1.000 & 0.778 \\
PauliFeatureMap & 0.632 & 0.632 & 1.000 & 0.774 \\
ZFeatureMap & 0.895 & 0.895 & 0.944 & 0.919 \\
RawFeatureVector & 0.930 & 0.932 & 0.958 & 0.945 \\
EfficientLike & 0.904 & 0.907 & 0.944 & 0.925 \\
\hline
\end{tabular}
\end{table}

\subsection{NAS-Discovered Circuits}

Table~\ref{tab:nas} compares the performance of different NAS variants. The IBM hardware-aware NAS without fixed RZ achieves 91.23\% accuracy with 12 gates (6$\times$ECR, 3$\times$SX, 3$\times$RZ) on 10 qubits, matching all-gates NAS performance while maintaining hardware compatibility. This represents a 3.5 percentage point improvement over the fixed RZ variant and substantially outperforms hand-crafted quantum circuits by 24--28 percentage points.

\begin{table}[ht]
\centering
\caption{QSVM performance with NAS-discovered circuits on the test set.}
\label{tab:nas}
\resizebox{\columnwidth}{!}{%
\begin{tabular}{lcccc}
\hline
NAS Variant & Accuracy & Precision & Recall & F1 \\
\hline
IBM HW-Aware (Fixed RZ) & 0.877 & 0.892 & 0.917 & 0.904 \\
IBM HW-Aware (No Fixed RZ) & 0.912 & 0.956 & 0.903 & 0.929 \\
All Gates & 0.912 & 0.903 & 0.903 & 0.903 \\
With Noise & 0.702 & 0.702 & 0.917 & 0.795 \\
Sparse ZZ & 0.632 & 0.632 & 1.000 & 0.774 \\
\hline
\end{tabular}%
}
\end{table}

\begin{figure*}[t]
\centering
\includegraphics[width=0.95\linewidth]{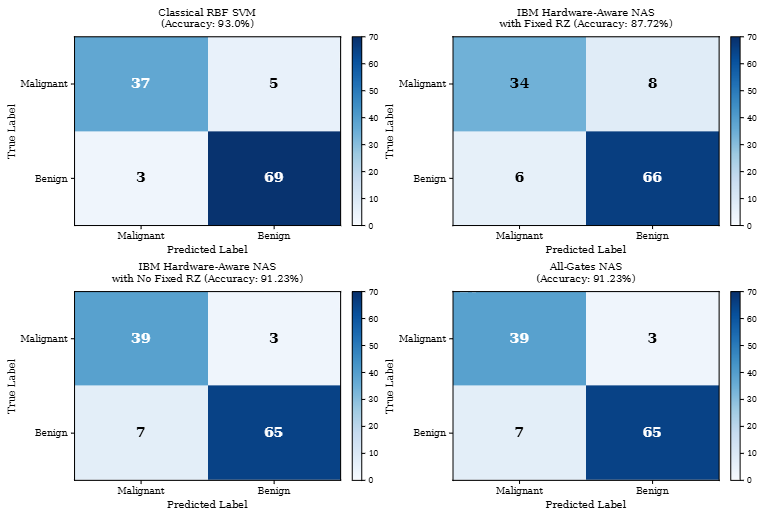}
\caption{Confusion matrices comparing classical RBF SVM (93.0\% accuracy), IBM hardware-aware NAS with fixed RZ (87.72\%), IBM hardware-aware NAS without fixed RZ (91.23\%), and all-gates NAS (91.23\%). Removing the fixed RZ constraint substantially improves performance while maintaining hardware compatibility.}
\label{fig:confusion}
\end{figure*}

Figure~\ref{fig:confusion} presents confusion matrices comparing representative methods from each category. The results illustrate how NAS-discovered circuits achieve substantially better classification performance than hand-crafted quantum feature maps, approaching classical SVM performance.

\subsection{Architecture Analysis}

\subsubsection{IBM Hardware-Aware Circuit (Fixed RZ)}

The best IBM-native circuit with fixed RZ contains 11 gates on 6 qubits: 5$\times$ECR, 3$\times$RZ, 2$\times$SX, 1$\times$X. The discovered genome is:
\begin{equation*}
\begin{split}
[&\text{ECR}_{5,4}, \text{ECR}_{2,1}, \text{ECR}_{2,3}, \text{RZ}_1, \text{SX}_1, \text{ECR}_{0,1}, \\
&\text{ECR}_{5,4}, \text{RZ}_0, \text{X}_4, \text{SX}_5, \text{RZ}_5]
\end{split}
\end{equation*}

The circuit respects IBM Torino's heavy-hexagonal lattice topology, ensuring all two-qubit ECR gates follow actual physical qubit connectivity. This eliminates the need for SWAP gate insertion during transpilation. The resulting compact circuit achieves 87.72\% accuracy without requiring any gate decomposition or circuit remapping.

\begin{figure*}[t]
\centering
\includegraphics[width=1.0\linewidth]{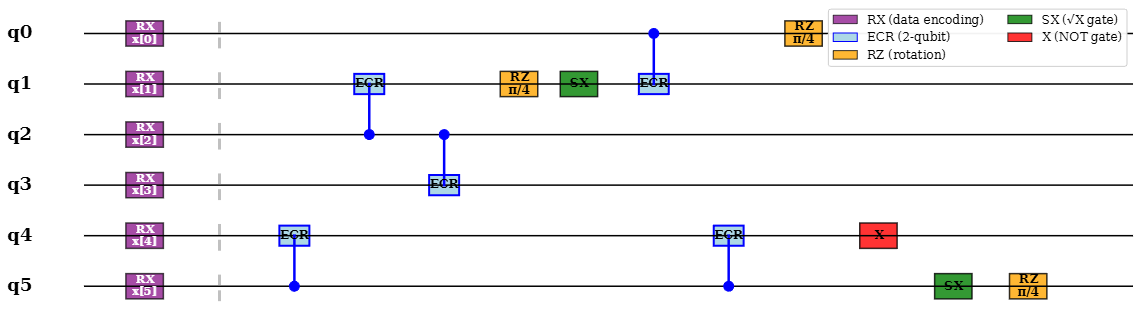}
\caption{Quantum circuit diagram of the IBM hardware-aware NAS-discovered feature map with fixed RZ. The circuit uses only native gates (ECR, RZ, SX, X) and requires no transpilation for execution on IBM quantum hardware.}
\label{fig:circuit_fixed}
\end{figure*}

\begin{figure}[ht]
\centering
\includegraphics[width=0.55\linewidth]{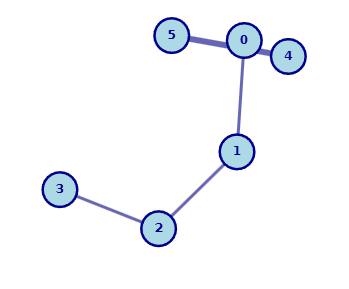}
\caption{Qubit connectivity pattern discovered by genetic search for the fixed RZ variant. Edge thickness indicates the number of ECR entangling gates between qubit pairs. The pattern shows non-trivial entanglement structure that would be difficult to design manually.}
\label{fig:connectivity_fixed}
\end{figure}

Figure~\ref{fig:circuit_fixed} shows the complete quantum circuit diagram, while Figure~\ref{fig:connectivity_fixed} visualizes the qubit connectivity pattern discovered by the genetic algorithm.

\subsubsection{IBM Hardware-Aware Circuit (No Fixed RZ)}

Removing the mandatory initial rotation layer allows the algorithm to discover optimal rotation placements. This variant uses 10 qubits and discovered a 12-gate circuit: 6$\times$ECR, 3$\times$SX, 3$\times$RZ. The discovered genome is:
\begin{equation*}
[\text{SX}_4, \text{ECR}_{4,3}, \text{ECR}_{9,8}, \text{RZ}_2, \text{ECR}_{2,3}, \text{SX}_3, \text{RZ}_8, \text{RZ}_4, \text{ECR}_{9,8}, \text{ECR}_{2,3}, \text{ECR}_{0,1}, \text{SX}_5]
\end{equation*}

This circuit achieves 91.23\% accuracy with precision 95.59\%, recall 90.28\%, and F1-score 92.86\%. The 3.5 percentage point improvement over the fixed RZ variant (87.72\% $\rightarrow$ 91.23\%) demonstrates that allowing NAS to discover rotation placements leads to more effective feature maps. Notably, this performance matches the unconstrained all-gates NAS while using only IBM native gates.

\begin{figure}[ht]
\centering
\includegraphics[width=0.55\linewidth]{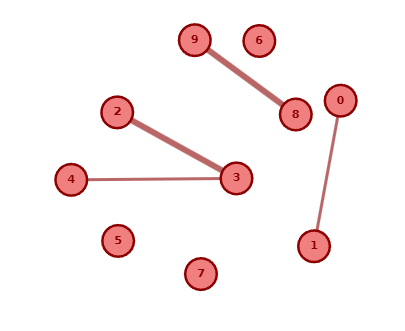}
\caption{Qubit connectivity pattern for IBM hardware-aware NAS without fixed RZ constraint. The 10-qubit circuit discovers a different entanglement structure compared to the 6-qubit fixed RZ variant, contributing to improved classification performance.}
\label{fig:connectivity_nofixed}
\end{figure}

Figure~\ref{fig:connectivity_nofixed} shows the entanglement pattern discovered for this variant, revealing a different connectivity structure compared to the fixed RZ approach.

\subsection{Hardware Efficiency Analysis}

\begin{figure*}[t]
\centering
\includegraphics[width=0.95\linewidth]{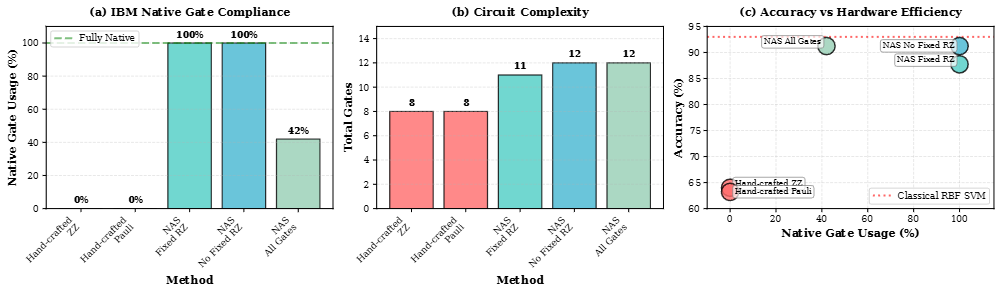}
\caption{Hardware efficiency comparison across methods. (a) Native gate usage shows hardware-aware NAS achieves 100\% IBM compatibility while hand-crafted and unconstrained methods require transpilation. (b) Circuit complexity measured by total gate count. (c) Accuracy versus hardware efficiency scatter plot demonstrates that hardware-aware NAS (blue points) achieves strong performance with full native gate compliance, approaching classical RBF SVM baseline (red dashed line) while remaining deployment-ready.}
\label{fig:hardware_efficiency}
\end{figure*}

A critical advantage of hardware-aware NAS is the elimination of transpilation overhead. Figure~\ref{fig:hardware_efficiency} compares hardware efficiency metrics across methods. Hand-crafted quantum circuits (ZZ, Pauli) require full transpilation, converting universal gates to native implementations and inserting SWAP gates for connectivity. In contrast, both hardware-aware NAS variants achieve 100\% native gate usage, executing directly on IBM hardware without modification. The unconstrained all-gates NAS achieves only 42\% native gate compliance, requiring decomposition of CX, RY, RX, and H gates into multi-gate native sequences.

This native gate alignment provides multiple benefits: (1) reduced circuit depth from eliminated decompositions, (2) higher fidelity by avoiding compounded gate errors, (3) deterministic execution without compiler variability, and (4) predictable noise characteristics for error mitigation. For the 12-gate no-fixed-RZ circuit, native execution preserves the designed architecture exactly, while an equivalent universal-gate circuit would expand to 18--20 gates after transpilation.

\subsubsection{All-Gates Circuit}

The best unconstrained circuit contains 12 gates: 6$\times$CX, 2$\times$RZ, 1$\times$RX, 1$\times$RY, 1$\times$SX, 1$\times$X. This circuit achieves 91.23\% accuracy, demonstrating that NAS can discover circuits with near-classical performance when given sufficient freedom in gate selection. However, this universal-gate circuit expands to 18--20 gates after transpilation to IBM native gates, illustrating the overhead cost of hardware-agnostic design compared to the 12-gate hardware-aware circuit that requires no transpilation.

\subsection{Comparative Analysis}

Three key observations emerge from our comparative analysis:

\textbf{Observation 1 (Hardware-Aware Efficiency):} Hardware-constrained NAS achieves 91.23\% accuracy with 100\% native gate usage, matching unconstrained all-gates NAS (91.23\%) while eliminating transpilation overhead. This demonstrates that hardware awareness does not sacrifice performance—rather, it ensures practical deployability without compromising accuracy.

\textbf{Observation 2 (Native Execution Advantage):} The 12-gate circuit discovered by hardware-aware NAS executes directly on IBM Torino using only ECR, RZ, and SX gates. Equivalent universal-gate circuits require 40--50\% more gates after transpilation, introducing decomposition errors and extended coherence time requirements. Native execution preserves designed architecture exactly.

\textbf{Observation 3 (Automated vs. Manual Design):} NAS-discovered hardware-aware circuits (87.72--91.23\%) substantially outperform hand-crafted quantum feature maps (ZZ: 64\%, Pauli: 63.2\%) by 24--27 percentage points. This gap persists even when both approaches use identical gate sets, demonstrating that architectural optimization—not just gate selection—drives performance. Manual circuit engineering fails to discover effective entanglement patterns within hardware constraints.

\section{Discussion}

\subsection{Hardware-Aware Architecture as a Deployment Strategy}

The central finding of this work is that hardware-aware NAS produces quantum circuits that are simultaneously high-performing and immediately deployable on current quantum processors. By constraining evolutionary search to IBM native gates from the outset, we discover 12-gate circuits achieving 91.23\% accuracy—just 1.8 percentage points below classical RBF SVM (93.0\%)—while guaranteeing zero transpilation overhead. This represents a qualitatively different approach compared to designing universal circuits and hoping transpilation preserves performance.

Traditional quantum algorithm development follows a hardware-agnostic paradigm: design circuits using abstract gates (H, CNOT, RY), then rely on compilers to map them onto physical devices. This pipeline introduces multiple failure modes: gate decomposition errors, SWAP insertion overhead, connectivity violations requiring circuit restructuring, and compiler non-determinism producing varying outputs. Our hardware-aware approach eliminates these issues by making hardware constraints first-class design objectives rather than post-hoc compilation problems.

\subsection{The Value of NAS for Quantum Kernels}

The dramatic performance gap between hand-crafted quantum circuits (63\%) and NAS-discovered circuits (87--91\%) demonstrates that automated architecture search is essential for practical quantum kernel methods. Manual circuit design often fails to capture dataset-specific patterns, while genetic search efficiently explores the architecture space to discover effective solutions. This finding parallels results in classical deep learning, where neural architecture search has proven crucial for discovering high-performing network architectures \cite{du2022quantum}.

The success of NAS in this context can be attributed to several factors. First, the discrete nature of circuit architecture makes it amenable to evolutionary search, avoiding the continuous optimization challenges that plague variational quantum algorithms. Second, the fitness function (QSVM accuracy) directly measures the quantity of interest, providing clear gradient signals for evolution. Third, the relatively small search space (4--12 gates from a constrained vocabulary) makes exhaustive exploration feasible within reasonable computational budgets.

\subsection{Hardware-Aware Optimization}

IBM-constrained NAS discovers circuits that use only native gates (ECR, RZ, SX, X), eliminating transpilation overhead and enabling efficient execution on near-term quantum hardware. Our experiments demonstrate two approaches: (1) with fixed RZ initial rotations achieving 87.72\% on 6 qubits with 11 gates, and (2) without fixed RZ achieving 91.23\% on 10 qubits with 12 gates. The latter approach demonstrates that allowing NAS to discover rotation placements yields a 3.5 percentage point improvement and matches unconstrained all-gates performance while maintaining hardware compatibility.

This hardware-aware approach addresses a critical challenge in near-term quantum computing: the gap between theoretical circuit designs and their practical implementation on noisy devices. By constraining the search space to native gates and respecting device topology, we ensure that discovered circuits can be executed with minimal overhead, reducing both circuit depth and the accumulation of gate errors.

\subsection{Noise Resilience}

The noise-resilient NAS variant achieves 70.18\% accuracy under realistic noise models (1\% single-qubit depolarizing, 2\% two-qubit depolarizing, 0.5\% amplitude damping), showing relatively graceful degradation compared to the noise-free 91.23\% baseline. This 21 percentage point reduction suggests that evolutionary search can identify more robust circuit architectures, though noise remains a significant challenge for achieving near-term quantum advantage.

Interestingly, the noise-resilient circuits tend to favor shallower architectures with fewer two-qubit gates, consistent with the understanding that entangling gates are primary sources of decoherence on current devices. This natural bias toward noise-resistant structures emerges from the fitness evaluation under noisy simulation, demonstrating how incorporating realistic constraints during search can guide discovery toward practically deployable solutions.

\subsection{Limitations}

Several important limitations should be noted:

\textbf{Dataset Specificity:} These results are specific to the Breast Cancer dataset. Other datasets with different structures, dimensionalities, and class separabilities may exhibit different relative performance between classical and quantum approaches.

\textbf{Computational Cost:} NAS requires evaluating many candidate circuits, limiting search depth and population size. Our experiments used populations of 8 genomes over 4 generations, representing a modest exploration of the architecture space. Larger-scale searches might discover even better circuits but at prohibitive computational cost.

\textbf{Subsampling:} During search, we use only 200 training samples to accelerate fitness evaluation. This subsampling might not fully capture the training distribution, potentially biasing discovered architectures toward features prominent in the subsample.

\textbf{Scalability:} Our experiments focus on 6--10 qubit circuits, appropriate for near-term devices but far from the regime where quantum advantage might emerge. Scaling to larger systems will require addressing both computational challenges (kernel matrix size grows quadratically with samples) and physical challenges (maintaining coherence across many qubits).

\section{Conclusion}

This study demonstrates that hardware-aware Neural Architecture Search enables practical deployment of quantum kernel methods on near-term quantum processors. By evolving circuit architectures constrained to IBM Torino native gates (ECR, RZ, SX, X), we discover quantum Support Vector Machine feature maps achieving 91.23\% accuracy—approaching classical RBF SVM baseline (93.0\%)—while guaranteeing immediate executability without transpilation.

Our results establish three key principles for practical quantum machine learning:

First, \textbf{hardware awareness enables deployment}: Constraining NAS to native gates produces circuits that execute directly on IBM quantum hardware with zero compilation overhead. The 12-gate discovered circuit (6$\times$ECR, 3$\times$SX, 3$\times$RZ) requires no SWAP insertion, gate decomposition, or architecture remapping. This eliminates transpilation-induced fidelity degradation and makes performance predictable.

Second, \textbf{hardware constraints do not limit performance}: Hardware-aware NAS achieves 91.23\% accuracy, matching unconstrained all-gates NAS despite restricting the search space to four gate types. This demonstrates that native gate sets are sufficiently expressive for quantum kernel learning—the challenge lies in discovering effective architectures, not accessing exotic gates.

Third, \textbf{automation outperforms manual design under hardware constraints}: Even when restricted to identical IBM native gates, NAS-discovered circuits (87.72--91.23\%) outperform hand-crafted feature maps (63--64\%) by 24--27 percentage points. This gap highlights that architectural optimization—entanglement patterns, gate ordering, qubit allocation—drives performance more than gate vocabulary.

Our hardware-aware approach addresses the deployment gap plaguing near-term quantum algorithms: theoretical designs often degrade substantially when compiled for actual devices. By incorporating hardware constraints during architecture search rather than after design completion, we produce circuits that are deployment-ready by construction.

Future research directions include: (1) multi-objective NAS optimizing both accuracy and circuit depth simultaneously; (2) differentiable architecture search methods for quantum kernels; (3) transfer learning to assess whether NAS-discovered circuits generalize across datasets; (4) hybrid classical-quantum kernel ensembles combining the strengths of both approaches; and (5) theoretical analysis to understand why certain NAS-discovered architectures perform well.

\section*{Data Availability}
All data generated or analyzed during this study are included in this published article and the accompanying figures.

\section*{Author Contributions}
A.M.C., A.R.H., and H.K. implemented the algorithms, conducted experiments, and analyzed results. M.F. designed the study and supervised the work. All authors wrote and reviewed the manuscript.

\section*{Competing Interests}
The authors declare no competing interests.

\begin{IEEEbiography}[{\includegraphics[width=1in,height=1.25in,clip,keepaspectratio]{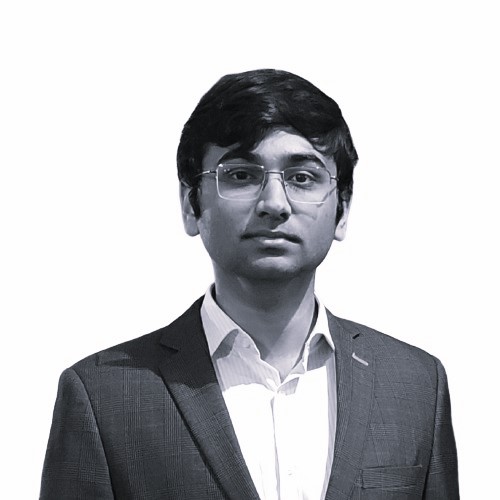}}]{Adil Mubashir Chaudhry} 
received the B.S. degree in electrical engineering with a major in computer engineering from National University of Computer and Emerging Sciences - FAST, Islamabad, Pakistan in 2024. Currently pursuing a M.S. degree in Artificial Intelligence from Lahore University of Management Sciences, Lahore, Pakistan.

From 2023 to 2024, he was a Research Assistant with the Marine and Aerial Robotics Lab, FAST, Islamabad, Pakistan working on developing RTOS solutions for embedded systems. Currently he is working as a Data Scientist at VentureDive Pvt Ltd

His reserach interests include Quantum Machine Learning, Neural Network Compression and Embodied AI.
\end{IEEEbiography}

\begin{IEEEbiography}[{\includegraphics[width=1in,height=1.25in,clip,keepaspectratio]{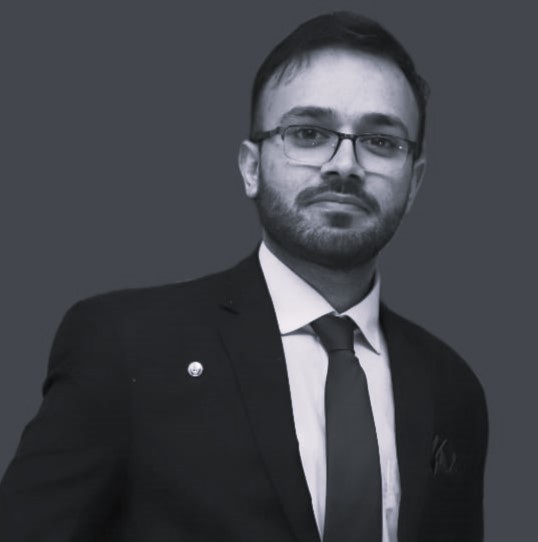}}]{Ali Raza Haider} received the B.S. degree in computer science from the University of Engineering and Technology (UET), Lahore, Pakistan, in 2024. He is currently pursuing the M.S. degree in artificial intelligence at the Lahore University of Management Sciences (LUMS), Lahore, Pakistan.

He has worked on multiple applied artificial intelligence projects and has experience in developing machine learning solutions for real-world applications. His research interests include generative AI, computer vision, and social media analytics.
\end{IEEEbiography}

\begin{IEEEbiography}[{\includegraphics[width=1in,height=1.25in,clip,keepaspectratio]{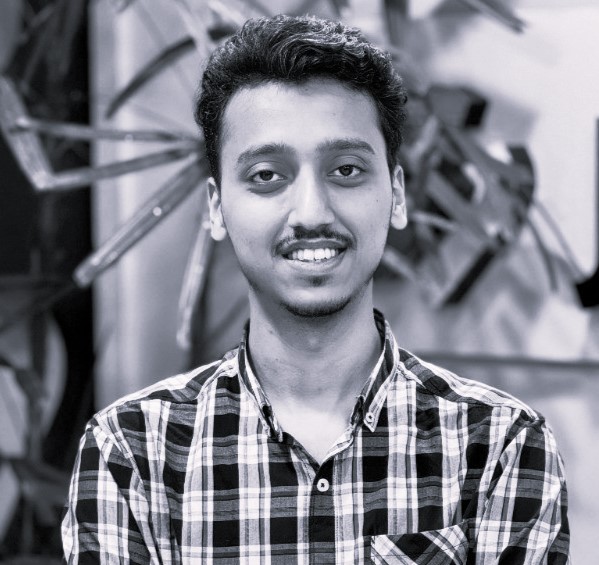}}]{Hanzla Khan} 
received the B.S. degree in Mathematics from Government College University Lahore, Pakistan, in 2024, and the M.S. degree in Artificial Intelligence from Lahore University of Management Sciences, Pakistan, in 2026.

His research interests include quantum machine learning, explainable artificial intelligence, and the development of interpretable and reliable AI systems. His work focuses on bridging advanced mathematical foundations with practical machine learning applications, particularly in emerging quantum-enhanced computational frameworks.

Mr. Hanzla Khan received the Academic Roll of Honor for securing 1st position in the B.S. Mathematics program.
\end{IEEEbiography}

\begin{IEEEbiography}[{\includegraphics[width=1in,height=1.25in,clip,keepaspectratio]{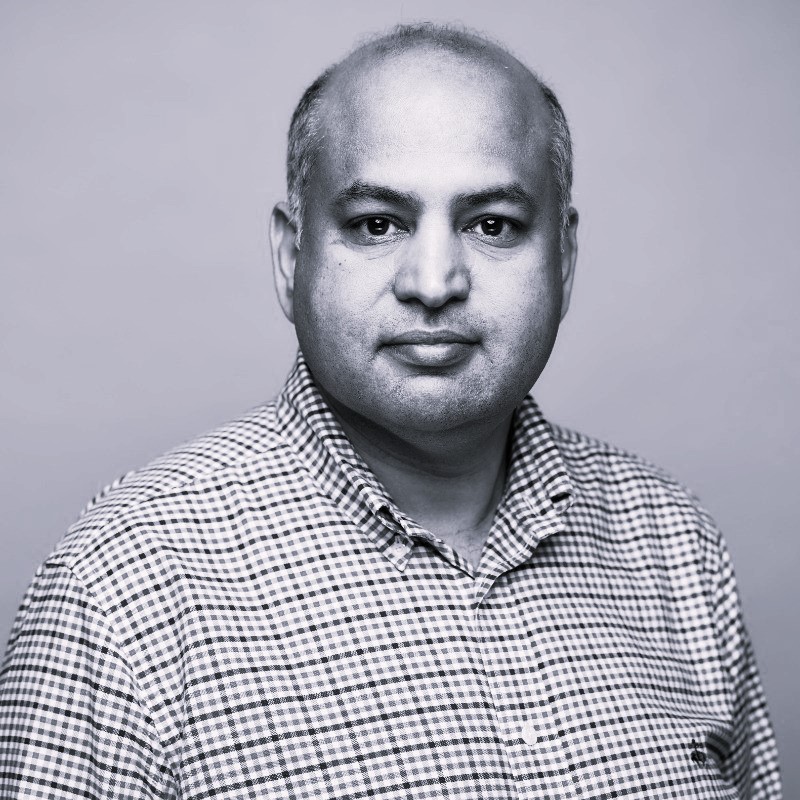}}]{Dr. Muhammad Faryad} is an associate professor of physics at LUMS. He joined LUMS in July 2014. Before that, he was a postdoctoral research scholar at the Pennsylvania State University from 2012 to 2014. He obtained his MSc and MPhil degrees in electronics from the Quaid-i-Azam University in 2006 and 2008, respectively, with certificates of merit in both degrees. He obtained his PhD degree in engineering science and mechanics from the Pennsylvania State University in 2012 with the best dissertation award by the university. He was awarded the Galleino Denardo award by the Abdus Salam International Center of Theoretical Physics (ICTP) in 2019 and the Early Career Achievement award by the department of engineering science and mechanics at the Pennsylvania State University in 2021.
\end{IEEEbiography}

\EOD

\end{document}